\documentclass[aps,twocolumn,floatfix,superscriptaddress,prl]{revtex4-2}
\usepackage{mathrsfs}
\usepackage[normalem]{ulem}
\usepackage{xcolor}
\usepackage{graphicx}           
\usepackage{dcolumn}            
\usepackage{bm}                 
\usepackage{hyperref}           
\usepackage[latin1]{inputenc}   
\usepackage{pstricks}           
\usepackage[normalem]{ulem}
\usepackage{amsmath,amssymb}
\usepackage{version}
\usepackage{epstopdf}
\usepackage{lipsum}
\DeclareGraphicsExtensions{.png,.eps,.ps}
\usepackage{ulem}
\usepackage{float}
\usepackage{booktabs}
\usepackage{subfigure}
\usepackage{graphicx}
\usepackage{dcolumn}
\usepackage{bm}
\usepackage{braket}             
\usepackage{footmisc}          
\usepackage{float}

\definecolor{RED}{RGB}{186, 16, 22}

\begin{document}

	\title{Ultrafast electron vortex produced by a grating made of light}
	
	\author{Zichen Li}\email{These authors contribute equally to this work.}\address{School of Physics, Zhejiang Key Laboratory of Micro-Nano Quantum Chips and Quantum Control, Zhejiang University, Hangzhou, 310058, China}
	\author{Hao Liang}\email{These authors contribute equally to this work.}\address{Max-Planck-Institut f\"ur Kernphysik, Heidelberg, 69117, Germany}
	\author{Yuan Gu}\address{School of Physics, Zhejiang Key Laboratory of Micro-Nano Quantum Chips and Quantum Control, Zhejiang University, Hangzhou, 310058, China}
	\author{Jiaye Zhang}\address{School of Physics, Zhejiang Key Laboratory of Micro-Nano Quantum Chips and Quantum Control, Zhejiang University, Hangzhou, 310058, China}
	\author{Aofan Lin}\address{School of Physics, Zhejiang Key Laboratory of Micro-Nano Quantum Chips and Quantum Control, Zhejiang University, Hangzhou, 310058, China}    
	\author{Juan Du}\address{School of Physics, Zhejiang Key Laboratory of Micro-Nano Quantum Chips and Quantum Control, Zhejiang University, Hangzhou, 310058, China}    
	\author{ Sina Jacob}\address{Institut f\"ur Kernphysik, Goethe-Universit\"at Frankfurt am Main, Frankfurt am Main 60438, Germany}         
	\author{Maksim Kunitski}\address{Institut f\"ur Kernphysik, Goethe-Universit\"at Frankfurt am Main, Frankfurt am Main 60438, Germany}  
	\author{Till Jahnke}\address{Max-Planck-Institut f\"ur Kernphysik, Heidelberg, 69117, Germany}
	\author{Sebastian Eckart}\address{Institut f\"ur Kernphysik, Goethe-Universit\"at Frankfurt am Main, Frankfurt am Main 60438, Germany}    
	\author{Reinhard D\"orner}\address{Institut f\"ur Kernphysik, Goethe-Universit\"at Frankfurt am Main, Frankfurt am Main 60438, Germany}         
	\author{Kang Lin}\email{klin@zju.edu.cn}\address{School of Physics, Zhejiang Key Laboratory of Micro-Nano Quantum Chips and Quantum Control, Zhejiang University, Hangzhou, 310058, China}       
	
	\date{\today}

	\begin{abstract}
		{
			The generation of vortex matter waves carrying quantized orbital angular momentum is challenging and relies heavily on the material nanofabrication methods due to their extremely small de-Broglie wavelengths. Here, we introduce an all-optical method for generating an electron vortex by diffraction through a grating made of light. We realize the orbital angular momentum transfer between free electrons and photons by stimulated Compton scattering. The transferred angular momentum quantum number can be freely tuned. The method can be generalized to a broad range of charged particles, neutral atoms, and molecules of diverse masses. Our results open up novel opportunities for applications in free electron lasers and ultrafast electron microscopy by utilizing the orbital angular momentum degree of freedom of free electrons.}
		
	\end{abstract}
	
	\maketitle
	
	A classical particle's orbital angular momentum (OAM) describes its rotational motion about a specific axis, which makes OAM reference-frame dependent in classical physics. In contrast, a quantum particle's OAM manifests in the form of phase of the particle's wave function that changes as a function of the angle around its quantization axis. In the case of a free quantum particle, this quantization axis corresponds to its propagation direction, and the OAM manifests in addition as a vortex-shaped wavefront of the particle.  As an example, photons can easily be shaped to carry OAMs by optical components \cite{allen_orbital_1992}, e.g., a spiral phase plate, resulting in a phase singularity in the beam center and a transverse vortex wavefront. The direct consequence of this phase singularity is a node in the center of a light beam or a focus. Optical vortices have wide applications in the field of quantum information \cite{mair_entanglement_2001,fickler_quantum_2012}, high-resolution imaging \cite{willig_sted_2006,fang_orbital_2020}, optical tweezers methods \cite{he_direct_1995,oneil_intrinsic_2002,paterson_controlled_2001} and ultrafast physics \cite{fang_ultrafast_2025,fang_structured_2024}, including strong-field ionization \cite{de_ninno_photoelectric_2020,fang_photoelectronic_2021,fang_probing_2022,schmiegelow_transfer_2016,begin_orbital_2025,li_attosecond_2026} and high-harmonic generation \cite{zurch_strong-field_2012,gariepy_creating_2014,kong_controlling_2017,rego_generation_2019,dorney_controlling_2019,gui_second-harmonic_2021,hancock_second-harmonic_2021,zuo_non-cascade_2026,xu_molecular_2025}. Substantially more challenging is the generation of vortices for massive particles. Due to their extremely small de-Broglie wavelengths, these require demanding nanofabrication methods. The pioneering work of using graphite films to stack a spiral phase plate for electrons initiated the field of vortex matter wave engineering \cite{uchida_generation_2010}. Neutrons can also be shaped in the same way by manufacturing such phase plates made of Aluminum alloy \cite{clark_controlling_2015}. Alternatively, particles diffracted at a fork-structured hologram can give rise to vortex beams in nonzero orders \cite{verbeeck_production_2010,luski_vortex_2021,mcmorran_electron_2011}. On the other hand, it has been shown that vortex matter waves can be obtained by laser-assisted near-field interactions \cite{vanacore_ultrafast_2019}. It remains that an appropriate material is needed to mediate the photon-electron interaction given their different dispersion relations. 
	
On the other hand, as already suggested by Kapitza and Dirac in 1933, an electron beam can be diffracted by a standing light wave, i.e., without the use of material gratings, termed as the Kapitza-Dirac effect nowadays\cite{kapitza_reflection_1933,batelaan_colloquium_2007}. This concept actually provides an all-optical way to manipulate matter waves and has been widely applied in electron, atomic, and molecular diffractions \cite{freimund_observation_2001,freimund_bragg_2002,gould_diffraction_1986,nairz_diffraction_2001,stickler_enantiomer_2021}, including cold atoms and Bose-Einstein condensates \cite{ovchinnikov_diffraction_1999}. In a seminal work in 2015, Handali \textit{et al.} proposed creating electron vortex beams using the Kapitza-Dirac effect via diffraction through two optical vortices intersecting at an angle of 45$^\circ$ \cite{handali_creating_2015}. A similar strategy was suggested by Koz$\acute{a}$k in 2021 \cite{kozak_electron_2021}. Until the present day, neither has been realized. Very recently, the conventional Kapitza-Dirac effect is extended to the ultrafast time domain by replacing the stationary standing wave with an ultrashort one made of two counterpropagating femtosecond laser pulses \cite{lin_ultrafast_2024}. This establishes a unique tool for manipulating matter-wave dynamics on ultrafast time scales.
	
	\begin{figure}
		\centering
		\includegraphics[width=1.0\columnwidth]{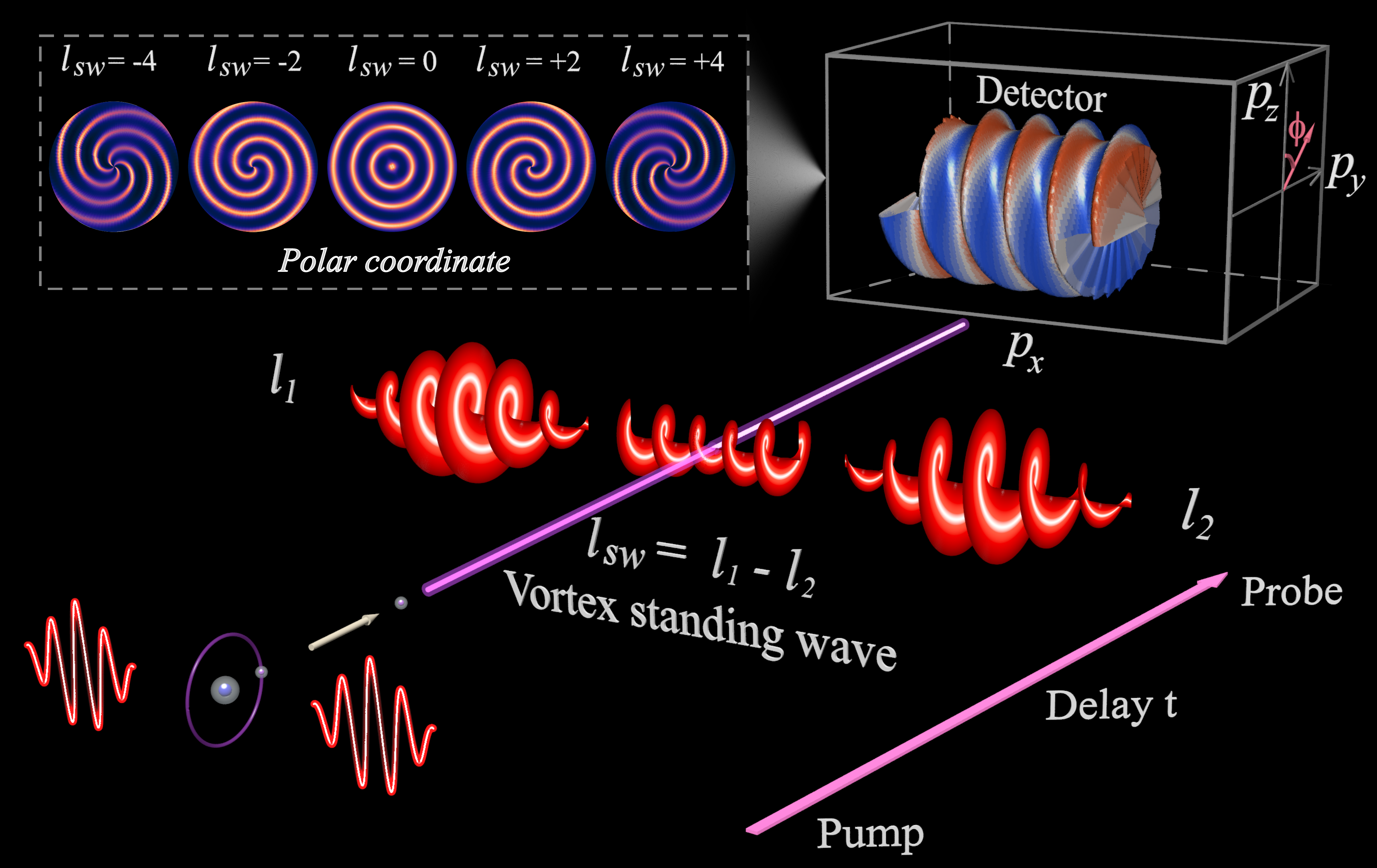}
		\caption{Schematic diagram of electron vortex generation. The electron wave packet is released from a xenon atom through strong-field ionization by a highly intense, femtosecond standing wave laser pulse (pump). After a time delay, a non-ionizing vortex standing light wave (probe) is applied to diffract the electron wave packet. The pump and probe pulses are both linearly polarized. The delay between the pump and probe allows the electron wave packet to spread and travel to the donut-shaped high-intensity region of the vortex probe pulse. The two arms of the vortex standing light wave carry OAMs with topological charges of $l_1$ and $l_2$, respectively. The total topological charge of the synthesized vortex standing light wave is $l_{sw}$ = $l_1$ - $l_2$. Three-dimensional electron momenta are recorded by the detector.\label{Fig1}}
	\end{figure}

	In this Letter, we demonstrate that the problematic requirement for auxiliary material in the generation of vortex matter waves can be bypassed by leveraging the recently discovered ultrafast Kapitza-Dirac effect \cite{lin_ultrafast_2024}. We achieve all-optical generation of free electron vortices in momentum space by exposing the electrons to a standing light wave grating made of two counterpropagating femtosecond optical vortices, as schematically illustrated in Fig. \ref{Fig1}. By independently controlling the topological charges $l_1$ and $l_2$ of the two optical vortices, we construct a vortex grating with an adjustable topological charge of $l_{sw}$ = $l_1$ - $l_2$. For the sake of simplicity, we consider two counterpropagating optical vortices traveling along the positive and negative $x$-direction with linear polarizations along the $z$-direction (atomic units are used unless stated otherwise):
	\begin{align}\label{Eq1and2}
		&E_{+}(\rho,x,\phi,t)=E_{0}(\rho)\cos(-kx+\omega t+l_{1}\phi)\\ 
		&E_{-}(\rho,x,\phi,t)=E_{0}(\rho)\cos(kx+\omega t+l_{2}\phi),    
	\end{align}
	where $E_{0}(\rho)$ is the amplitude of the electric field at a radial distance $\rho$ from the central axis, $\omega$ is the angular frequency, $\bm{k}$ is the wave vector, $\phi$ is the azimuthal angle in the plane perpendicular to $\bm{k}$. The synthesized vortex standing light wave is
	\begin{equation}\label{Eq3}
		E_{sw} = 2 E_{0}(\rho)\cos(\omega t + \frac{l_{1}+l_{2}}{2}\phi) \cos(kx - \frac{l_{1}-l_{2}}{2}\phi).
	\end{equation}
	As a result, the time-averaged potential of the vortex grating is
	\begin{equation}\label{Eq4}
		U_{sw} = \overline{{\left\lvert \frac{E_{sw}}{\omega}\right\rvert}^2} = U_{0}(\rho)\cos[2kx-(l_{1}-l_{2})\phi].
	\end{equation}
	From Eq. (\ref{Eq4}), we obtain the topological charge of the synthesized vortex grating $l_{sw} $= $l_1$ - $l_2$. Obviously, $l_{sw}$ vanishes when $l_1$ = $l_2$, which returns to the original ultrafast Kapitza-Dirac effect \cite{lin_ultrafast_2024}. In such a standing light wave, electrons can undergo stimulated Compton scattering by absorbing one photon from one direction, followed by stimulated emission to the opposite direction. As a result, the electrons receive a linear momentum kick of integer numbers of 2$\bm{k}$ (i.e., twice the photon momentum) in conjunction with a zero OAM transfer $l_1$ - $l_2$ = 0. Only linear momentum is exchanged between photons and electrons in this case, leading to observable time-dependent diffraction fringes in momentum space\cite{lin_ultrafast_2024}. This is different if the aforementioned vortex standing light wave is employed in case of $l_1 \neq l_2$. The linear momentum transfer is additionally accompanied by a stimulated OAM transfer of $l_{sw}$ = $l_1$ - $l_2$, resulting in a twisted electron vortex.

	We implemented the OAM transfer protocol in a pump-probe scheme shown in Fig. \ref{Fig1}, with details given in the Supplementary Material. An initial electron wave packet is released from a neutral xenon atom upon strong-field ionization using an intense pulsed femtosecond standing wave (pump, $\sim1.3\times 10^{14}$ W/cm$^2$). After a variable time delay, a weak, non-ionizing, femtosecond standing light wave (probe, $\sim3.3\times 10^{12}$ W/cm$^2$) is applied to diffract the ionized electron wave packet. The pump and probe pulses are generated by a commercial Yb:KGW laser system (Light Conversion PHAROS, 260 fs, 1030 nm, 20 W). The polarizations of the pump and probe pulses are set to circular or linear by adjusting the half- and quarter-wave plates. The probe pulse is shaped into an optical vortex by reflecting off a Spatial Light Modulator (SLM, UPOLabs HDSLM80Rplus) before being split into two counterpropagating pathways to form a vortex standing light wave. The ionizing pump standing light wave is created by two counterpropagating pulses without a spatial light modulator. The pump and probe pulses are focused into the same target volume of a xenon gas jet by two $f$ = 25 cm lenses. The ionization produced electrons and ions are measured in coincidence by employing a COLd Target Recoil Ion Momentum Spectroscopy (COLTRIMS) reaction microscope \cite{dorner_cold_2000}.
	
	\begin{figure}
		\centering
		\includegraphics[width=1.0\columnwidth]{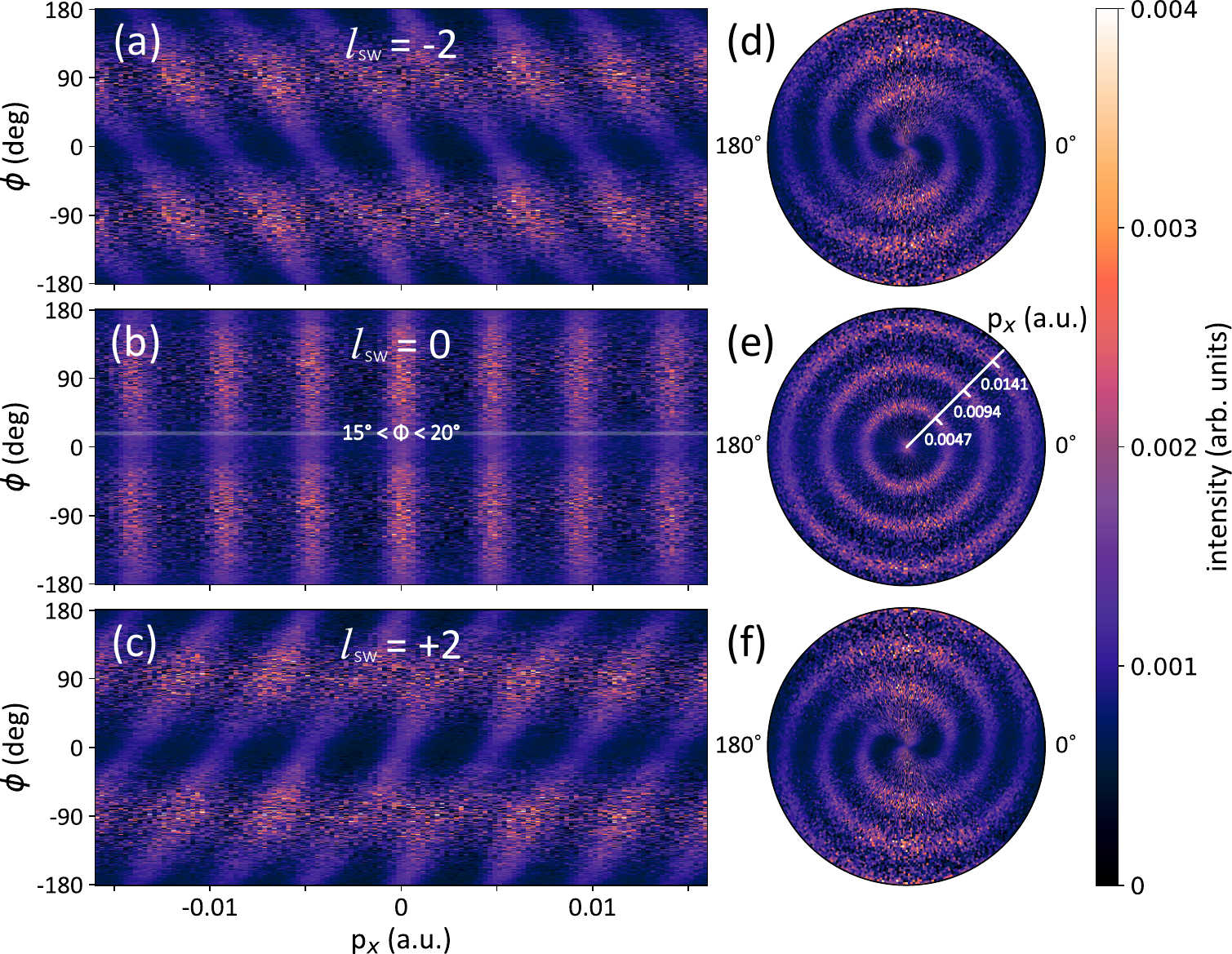}
		\caption{Measured momentum distributions of electron vortices driven by a linearly polarized laser field. The electron wave packet is created by strong-field ionization in a pulsed non-OAM standing light wave and diffracted at a second time delayed (delay of 50 ps) pulsed standing light wave with (a,c) or without (b) OAM. (a-c) Measured two-dimensional electron momentum distribution along the laser-propagation direction $p_x$ as a function of the azimuthal angle $\phi$ in the $y$-$z$ plane. Panels (a-c) show the data for topological charges of $l_{sw}$ = -2, 0, and 2, respectively. (d-f) Polar plots of (a-c) by taking $p_x$ as the radius. The datasets are normalized to each azimuthal angle $\phi$ for the purpose of eliminating the anisotropy in the initial momentum distribution. Noticeably, the blurred areas around $\phi$ = $\pm$ 90$^\circ$ in (a-c) result from the anisotropic distribution of the ionized electron wave packet released by a linearly polarized pump pulse, where the electron momentum extends along the laser polarization direction of $\phi$ = 0$^\circ$ and $\pm$ 180$^\circ$ due to the acceleration exerted by the laser electric field \cite{Keldysh_ionization_1965}. In addition, the condition $p_r = \sqrt{p^2_y + p^2_z} >$ 0.15 a.u. was applied to remove Coulomb-focused electrons \cite{brabec_coulomb_1996,blaga_strong-field_2009,gu_dynamical_2026}.\label{Fig2}}
	\end{figure}
	
	Figure \ref{Fig2} shows the results of our experiment. For reference, we depict the case of using a light grating with $l_{sw}$ = 0 (planar wavefront) in panel (b). It shows the measured electron momentum distribution along the laser-propagation direction $p_x$ as a function of the azimuthal angle $\phi$ in $y$-$z$ plane, as indicated in Fig. \ref{Fig1}. The vertical parallel diffraction fringes indicate that the phase of the electron wave packet is $\phi$-independent, which is the same as observed for the original ultrafast Kapitza-Dirac effect reported in ref. \cite{lin_ultrafast_2024}. Here, the fringe spacing of $\sim4.7\times 10^{-3}$ a.u. is determined by $\lambda_{sw}/2t$, where $\lambda_{sw}$ = 1030 nm is the driving wavelength of the standing light wave, and $t$ is the pump-probe delay of 50 ps\cite{lin_ultrafast_2024}. The reason for choosing such a relatively long delay is to allow the initial electron wave packet to spread in real space to cross the central dark region of the donut-shaped intensity profile of the probe standing wave. Such a profile is typical for optical vortices (for further details see the Supplementary Material). Figure \ref{Fig2}(a) shows the result for the topological charge of the probe pulse of  $l_{sw}$ = -2, where the diffraction fringes distinctly tilt as compared to the vertical ones in Fig. \ref{Fig2}(b). The tilt of the fringes shows that the phase of the kicked wave packet replica is twisted compared to the parent wave packet it interferes with. Figure \ref{Fig2}(c) shows the same distribution for $l_{sw}$ = 2, demonstrating the controllable helicity of the electron vortex. In addition, we find that regardless of whether we employed circular or linear polarization for the ionization (pump) and diffraction (probe) pulses, vortex electrons are reliably produced. The corresponding results for circularly polarized pulses are presented in the Supplementary Material. Figures \ref{Fig2}(d-f) show the same data as Figs. \ref{Fig2}(a-c) in a polar representation by taking $p_x$ as the radius. These polar plots serve as a direct indicator of the topological charge of the electron vortex. As a reference, the polar plot of diffraction by a planar grating exhibits a series of equally spaced concentric rings, indicating $l_{sw}$ = 0 in Fig. \ref{Fig2}(e). In comparison, the two electron vortices presented in Figs. \ref{Fig2}(d) and \ref{Fig2}(f) twist clockwise or counterclockwise with two-fold symmetries, indicating $l_{sw}$ = $\mp$2, respectively.
	
	\begin{figure}
		\centering
		\includegraphics[width=1.0\columnwidth]{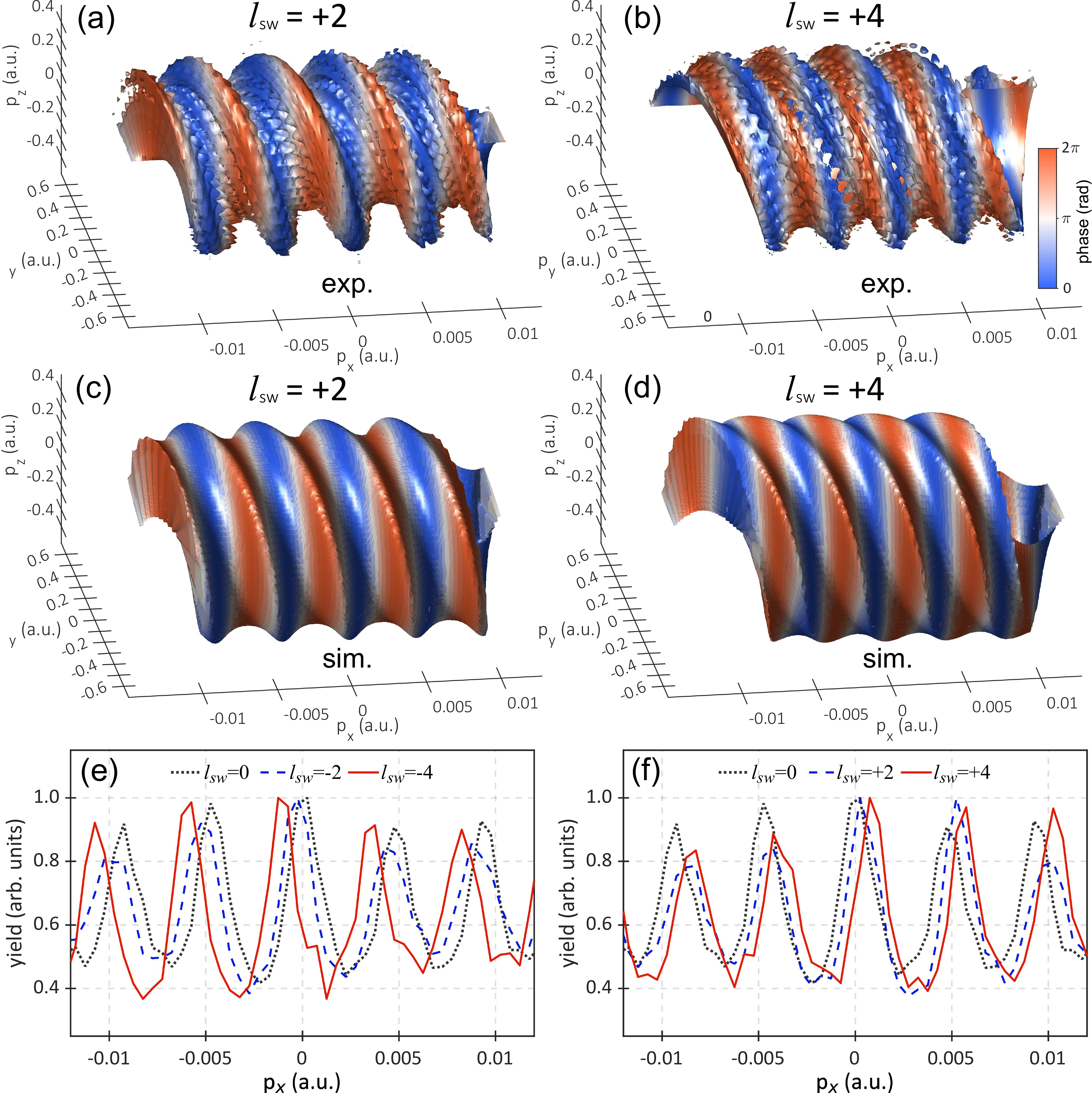}
		\caption{Three-dimensional momentum distributions of electron vortices. (a-d) Three-dimensional (3D) momentum distribution of the electron vortex with $l_{sw}$ = 2 and 4 obtained from the experimental data (panels a, b) and the simulation (panels c, d). Plotted are isosurfaces at 10\% of a normalized density. The data in blurred areas around $\phi$ = $\pm$ 90$^\circ$  is filled by those recorded around 0$^\circ$ and $\pm$ 180$^\circ$ according to the rotational symmetry. The color map artificially displays the phase change of the wavefront of the electron vortex. (e),(f) 1D momentum distribution along the laser-propagation direction by selecting $\phi \in$[15$^\circ$, 20$^\circ$] (shaded areas in Fig. \ref{Fig2} in the main text and Fig. S6 in the Supplementary Material). The black dotted, blue dashed, and red solid curves correspond to topological charge of $|l_{sw}|$ = 0, 2, and 4, respectively.\label{Fig3}}
	\end{figure}
	
	Figure \ref{Fig3}(a) shows the three-dimensional (3D) momentum distributions of the electron vortex of $l_{sw}$ = 2. We further generate high-order electron vortices by increasing the topological charge $l_{sw}$. Figure \ref{Fig3}(b) shows the reconstructed three-dimensional momentum distribution of the electron vortex with $l_{sw}$ = 4. The original electron momentum distributions for $l_{sw}$ = $\pm$4 for both linear and circular polarizations are presented in Supplementary Material. Intuitively, the diffraction fringes tilt more when the topological charge increases. Figures \ref{Fig3}(e),(f) show the one-dimensional momentum distribution along the light-propagation direction by selecting $\phi \in$[15$^\circ$, 20$^\circ$] to highlight the fringe tilt between high- and low-order electron vortices. The red solid and blue dashed curves are taken from the data of $\left\lvert l_{sw}\right\lvert$ = 4 and $\left\lvert l_{sw}\right\lvert$ = 2 for linear polarizations. As a reference, the black dotted curves display the momentum distribution for $l_{sw}$ = 0. This tilt is directly linked to the phase imprinted on the wavefront of the kicked electron wave packet. A comparison between red solid and blue dashed curves reveals that the fringe tilt is markedly larger for higher topological charges. Quantitatively, by fitting the peaks, we obtain the slope of the tilted diffraction fringes is $\frac{\partial p_x}{\partial \phi}$ = 3.2$\times 10^{-3}$ a.u./rad for $l_{sw}$ = 4, which is exactly two times larger compared to $l_{sw}$ = 2 of 1.6$\times 10^{-3}$ a.u./rad.

	\begin{figure}
		\centering
		\includegraphics[width=1.0\columnwidth]{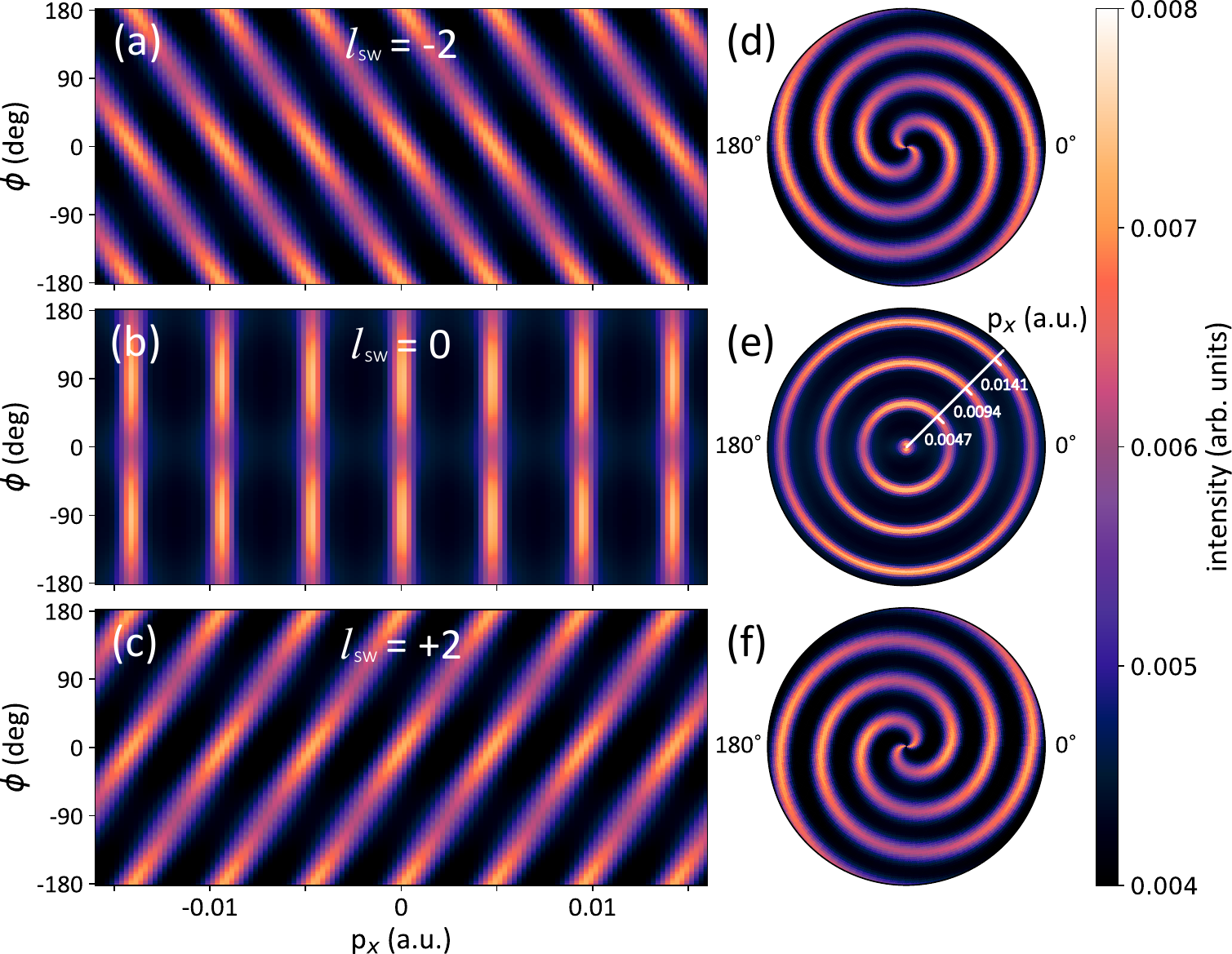}
		\caption{Simulated momentum distributions of electron vortices driven by a linearly polarized laser field. (a-c) Simulated two-dimensional electron momentum distribution along the laser-propagation direction $p_x$ as a function of the azimuthal angle $\phi$ in the $y$-$z$ plane. Panels (a-c) show the data for topological charges of $l_{sw}$ = -2, 0, and 2, respectively. (d-f) Polar plots of (a-c) by taking $p_x$ as the radius. The datasets are normalized to each azimuth angle $\phi$ for the purpose of eliminating the anisotropy in the initial momentum distribution.\label{Fig4}}
	\end{figure}
	
	In order to explain the original ultrafast Kapitza-Dirac effect, we proposed an interference picture to explain the time-dependent diffraction fringes. Importantly, in that case only integer multiples of the linear momenta of the photons are exchanged by stimulated Compton scattering \cite{lin_ultrafast_2024}. In our present study, using a vortex standing light wave, the underlying physics can be well understood by extending the interference picture with an additional OAM transfer between the free electrons and photons. Under field-free conditions, the parent electron wave packet accumulates a phase of $\varphi_0=-p^{2}_{x} t/2$. The phase of the replica of the wave packet created by stimulated backscattering of two vortex photons is modified to $\varphi_{\pm} = -\frac{(p_{x}\pm 2k)^{2}t}{2} \pm l_{sw} \times \phi$. The 2$\bm{k}$ and $l_{sw}\times \phi$ terms represent the aforementioned linear momentum and OAM transfer between the free electrons and photons, respectively. The parent wave packet and its replica interfere constructively when the phase difference is $\Delta\varphi = \varphi_{\pm} - \varphi_0$ = 2$n\pi$, n$\in\mathbb{Z}$. This results in peaks in the final electron momentum distribution at $(p_{x})_{n} \approx \pm \frac{n\pi}{kt}$ + $\frac{l_{sw}\phi}{2kt}$, resulting in the time-dependent and tilted diffraction fringes of slope $\frac{\partial p_x}{\partial \phi}$ = $\frac{l_{sw}}{2kt}$.

	Finally, we performed a numerical simulation by diffracting a 3D electron wave packet by the vortex standing light wave in a Laguerre-Gaussian mode. The initial electron wave packet in momentum space is given by the strong-field approximation \cite{milosevic_above-threshold_2006,becker_above-threshold_2002}. Subsequently, such an electron wave packet undergoes field-free evolution until being diffracted by a vortex grating (See Supplementary Material for the details of the simulation). Figures \ref{Fig3} and \ref{Fig4} show the simulation results, which agree well with the measurements. When the grating's topological charge vanishes, i.e., $l_{sw}$ = 0, the corresponding spectrum shows vertical fringes independent of the azimuthal angle. When $l_{sw}\neq$ 0, the diffraction fringes tilt and their slopes quantitatively satisfy $\frac{\partial p_x}{\partial \phi}$ = $\frac{l_{sw}}{2kt}$, as shown in Figs. \ref{Fig4}(a) and (c).

	In conclusion, we demonstrate an all-optical method for producing free electron vortices carrying quantized OAMs. By utilizing helical light structures instead of material phase plates or hologram masks, we eliminate the need for nanofabrication required by other methods for generating such vortices. The transferred OAM quantum numbers can, in principle, be tuned arbitrarily. This capability for the all-optical engineering of controllable vortex matter waves is easily generalized to a broad range of charged particles, neutral atoms, or molecules, with no fundamental limitation on mass. Our findings open the door to applications in vortex free electron laser generation and ultrafast electron microscopy by utilizing the OAM degree of freedom of free electrons.

	\begin{acknowledgments}
	K. L. acknowledges support by the National Natural Science Foundation of China (Project numbers 92576104, 12474348), the Natural Science Foundation of Zhejiang province (Project number LR26A040002), the Fundamental Research Funds for the Central Universities (Project number 226202500227), and the Startup Funding of Zhejiang University. R. D. acknowledges Funding by the European Union (ERC, Timing-FreePhase-Project number 101141762). Views and opinions expressed are however those of the author(s) only and do not necessarily reflect those of the European Union or the European Research Council. Neither the European Union nor the granting authority can be held responsible for them.
	\end{acknowledgments}

	 \bibliography{ref_prl_main}

@article{kozak_electron_2021,
	title = {Electron Vortex Beam Generation via Chiral Light-Induced Inelastic Ponderomotive Scattering},
	author = {Koz\'{a}k, M.},
	year = 2021,
	month = feb,
	journal = {ACS Photonics},
	volume = {8},
	number = {2},
	pages = {431--435},
	doi = {10.1021/acsphotonics.0c01650}
}

@article{uchida_generation_2010,
	title = {Generation of electron beams carrying orbital angular momentum},
	volume = {464},
	copyright = {http://www.springer.com/tdm},
	issn = {0028-0836, 1476-4687},
	url = {https://www.nature.com/articles/nature08904},
	doi = {10.1038/nature08904},
	language = {en},
	number = {7289},
	urldate = {2025-11-18},
	journal = {Nature},
	author = {Uchida, Masaya and Tonomura, Akira},
	month = apr,
	year = {2010},
	pages = {737--739},

}

@article{verbeeck_production_2010,
	title = {Production and application of electron vortex beams},
	volume = {467},
	copyright = {http://www.springer.com/tdm},
	issn = {0028-0836, 1476-4687},
	url = {https://www.nature.com/articles/nature09366},
	doi = {10.1038/nature09366},
	language = {en},
	number = {7313},
	urldate = {2025-11-18},
	journal = {Nature},
	author = {Verbeeck, J. and Tian, H. and Schattschneider, P.},
	month = sep,
	year = {2010},
	pages = {301--304},
}

@article{clark_controlling_2015,
	title = {Controlling neutron orbital angular momentum},
	volume = {525},
	issn = {0028-0836, 1476-4687},
	url = {https://www.nature.com/articles/nature15265},
	doi = {10.1038/nature15265},
	language = {en},
	number = {7570},
	urldate = {2025-11-18},
	journal = {Nature},
	author = {Clark, C. W. and Barankov, R. and Huber, M. G. and Arif, M. and Cory, D. G. and Pushin, D. A.},
	month = sep,
	year = {2015},
	pages = {504--506},
}

@article{vanacore_ultrafast_2019,
	title = {Ultrafast generation and control of an electron vortex beam via chiral plasmonic near fields},
	volume = {18},
	issn = {1476-1122, 1476-4660},
	url = {https://www.nature.com/articles/s41563-019-0336-1},
	doi = {10.1038/s41563-019-0336-1},
	language = {en},
	number = {6},
	urldate = {2025-11-18},
	journal = {Nat. Mater.},
	author = {Vanacore, G. M. and Berruto, G. and Madan, I. and Pomarico, E. and Biagioni, P. and Lamb, R. J. and McGrouther, 
	D. and Reinhardt, O. and Kaminer, I. and Barwick, B. and Larocque, H. and Grillo, V. and Karimi, E. and Garc{\'i}a De Abajo, F. J. and Carbone, F.},
	month = jun,
	year = {2019},
	pages = {573--579},
}

@article{mcmorran_electron_2011,
	title = {Electron Vortex Beams with High Quanta of Orbital Angular Momentum},
	volume = {331},
	issn = {0036-8075, 1095-9203},
	url = {https://www.science.org/doi/10.1126/science.1198804},
	doi = {10.1126/science.1198804},
	language = {en},
	number = {6014},
	urldate = {2025-11-18},
	journal = {Science},
	author = {McMorran, B. J. and Agrawal, A. and Anderson, I. M. and Herzing, A. A. and Lezec, H. J. and McClelland, J. J. and Unguris, J.},
	month = jan,
	year = {2011},
	pages = {192--195},
}

@article{luski_vortex_2021,
	title = {Vortex beams of atoms and molecules},
	volume = {373},
	issn = {0036-8075, 1095-9203},
	url = {https://www.science.org/doi/10.1126/science.abj2451},
	doi = {10.1126/science.abj2451},
	language = {en},
	number = {6559},
	urldate = {2025-11-18},
	journal = {Science},
	author = {Luski, A. and Segev, Y. and David, R. and Bitton, O. and Nadler, H. and Barnea, A. R. and Gorlach, A. and Cheshnovsky, O. and Kaminer, I. and Narevicius, E.},
	month = sep,
	year = {2021},
	pages = {1105--1109},
}

@article{lin_ultrafast_2024,
	title = {Ultrafast Kapitza-Dirac effect},
	volume = {383},
	issn = {0036-8075, 1095-9203},
	url = {https://www.science.org/doi/10.1126/science.adn1555},
	doi = {10.1126/science.adn1555},
	language = {en},
	number = {6690},
	urldate = {2025-11-18},
	journal = {Science},
	author = {Lin, Kang and Eckart, Sebastian and Liang, Hao and Hartung, Alexander and Jacob, Sina and Ji, Qinying and Schmidt, Lothar Ph. H. and Sch{\"o}ffler, Markus S. and Jahnke, Till and Kunitski, Maksim and D{\"o}rner, Reinhard},
	month = mar,
	year = {2024},
	pages = {1467--1470},
}

@article{fang_structured_2024,
	title = {Structured electrons with chiral mass and charge},
	volume = {385},
	issn = {0036-8075, 1095-9203},
	url = {https://www.science.org/doi/10.1126/science.adp9143},
	doi = {10.1126/science.adp9143},
	language = {en},
	number = {6705},
	urldate = {2025-11-18},
	journal = {Science},
	author = {Fang, Yiqi and Kuttruff, Joel and Nabben, David and Baum, Peter},
	month = jul,
	year = {2024},
	pages = {183--187},
}

@article{dorner_cold_2000,
	title = {Cold Target Recoil Ion Momentum Spectroscopy: a ‘momentum microscope’ to view atomic collision dynamics},
	volume = {330},
	issn = {03701573},
	shorttitle = {Cold {Target} {Recoil} {Ion} {Momentum} {Spectroscopy}},
	url = {https://linkinghub.elsevier.com/retrieve/pii/S037015739900109X},
	doi = {10.1016/S0370-1573(99)00109-X},
	language = {en},
	number = {2-3},
	urldate = {2025-11-18},
	journal = {Physics Reports},
	author = {D{\"o}rner, R. and Mergel, V. and Jagutzki, O. and Spielberger, L. and Ullrich, J. and Moshammer, R. and Schmidt-B{\"o}cking, H.},
	month = jun,
	year = {2000},
	pages = {95--192},
}

@article{batelaan_colloquium_2007,
	title = {Colloquium: Illuminating the Kapitza-Dirac Effect with Electron Matter Optics},
	volume = {79},
	doi = {10.1103/RevModPhys.79.929},
	number = {3},
	journal = {Rev. Mod. Phys.},
	publisher = {American Physical Society},
	author = {Batelaan, H.},
	month = jul,
	year = {2007},
	pages = {929--941},
}

@article{freimund_observation_2001,
	title = {Observation of the Kapitza–Dirac Effect},
	volume = {413},
	issn = {1476-4687},
	doi = {10.1038/35093065},
	number = {6852},
	journal = {Nature},
	author = {Freimund, Daniel L. and Aflatooni, Kayvan and Batelaan, Herman},
	month = sep,
	year = {2001},
	pages = {142--143},
}

@article{kapitza_reflection_1933,
	title = {The Reflection of Electrons from Standing Light Waves},
	volume = {29},
	issn = {0305-0041},
	doi = {10.1017/S0305004100011105},
	number = {2},
	journal = {Math. Proc. Cambridge Philos. Soc.},
	publisher = {Cambridge University Press},
	author = {Kapitza, P. L. and Dirac, P. A. M.},
	year = {1933},
	pages = {297--300},
}

@article{handali_creating_2015,
	title = {Creating electron vortex beams with light},
	volume = {23},
	copyright = {https://doi.org/10.1364/OA\_License\_v1\#VOR-OA},
	issn = {1094-4087},
	url = {https://opg.optica.org/abstract.cfm?URI=oe-23-4-5236},
	doi = {10.1364/OE.23.005236},
	language = {en},
	number = {4},
	urldate = {2025-12-03},
	journal = {Opt. Express},
	author = {Handali, Jonathan and Shakya, Pratistha and Barwick, Brett},
	month = feb,
	year = {2015},
	pages = {5236},
}

@article{nairz_diffraction_2001,
	title = {Diffraction of Complex Molecules by Structures Made of Light},
	volume = {87},
	issn = {0031-9007, 1079-7114},
	url = {https://link.aps.org/doi/10.1103/PhysRevLett.87.160401},
	doi = {10.1103/PhysRevLett.87.160401},
	number = {16},
	urldate = {2025-12-03},
	journal = {Phys. Rev. Lett.},
	author = {Nairz, Olaf and Brezger, Björn and Arndt, Markus and Zeilinger, Anton},
	month = sep,
	year = {2001},
	pages = {160401},
}

@article{gould_diffraction_1986,
	title = {Diffraction of atoms by light: The near-resonant Kapitza-Dirac effect},
	volume = {56},
	issn = {0031-9007},
	url = {https://link.aps.org/doi/10.1103/PhysRevLett.56.827},
	doi = {10.1103/PhysRevLett.56.827},
	number = {8},
	urldate = {2025-12-03},
	journal = {Phys. Rev. Lett.},
	author = {Gould, Phillip L. and Ruff, George A. and Pritchard, David E.},
	month = feb,
	year = {1986},
	pages = {827--830},
}

@article{stickler_enantiomer_2021,
	title = {Enantiomer Superpositions from Matter-Wave Interference of Chiral Molecules},
	volume = {11},
	issn = {2160-3308},
	url = {https://link.aps.org/doi/10.1103/PhysRevX.11.031056},
	doi = {10.1103/PhysRevX.11.031056},
	number = {3},
	urldate = {2025-12-04},
	journal = {Phys. Rev. X},
	author = {Stickler, Benjamin A. and Diekmann, Mira and Berger, Robert and Wang, Daqing},
	month = sep,
	year = {2021},
	pages = {031056},
}

@article{oneil_intrinsic_2002,
	title = {Intrinsic and Extrinsic Nature of the Orbital Angular Momentum of a Light Beam},
	volume = {88},
	issn = {0031-9007, 1079-7114},
	url = {https://link.aps.org/doi/10.1103/PhysRevLett.88.053601},
	doi = {10.1103/PhysRevLett.88.053601},
	language = {en},
	number = {5},
	urldate = {2025-12-04},
	journal = {Phys. Rev. Lett.},
	author = {O'Neil, A. T. and MacVicar, I. and Allen, L. and Padgett, M. J.},
	month = jan,
	year = {2002},
	pages = {053601},
}

@article{he_direct_1995,
	title = {Direct Observation of Transfer of Angular Momentum to Absorptive Particles from a Laser Beam with a Phase Singularity},
	volume = {75},
	copyright = {http://link.aps.org/licenses/aps-default-license},
	issn = {0031-9007, 1079-7114},
	url = {https://link.aps.org/doi/10.1103/PhysRevLett.75.826},
	doi = {10.1103/PhysRevLett.75.826},
	language = {en},
	number = {5},
	urldate = {2025-12-04},
	journal = {Phys. Rev. Lett.},
	author = {He, H. and Friese, M. E. J. and Heckenberg, N. R. and Rubinsztein-Dunlop, H.},
	month = jul,
	year = {1995},
	pages = {826--829},
}

@article{paterson_controlled_2001,
	title = {Controlled Rotation of Optically Trapped Microscopic Particles},
	volume = {292},
	issn = {0036-8075, 1095-9203},
	url = {https://www.science.org/doi/10.1126/science.1058591},
	doi = {10.1126/science.1058591},
	language = {en},
	number = {5518},
	urldate = {2025-12-04},
	journal = {Science},
	author = {Paterson, L. and MacDonald, M. P. and Arlt, J. and Sibbett, W. and Bryant, P. E. and Dholakia, K.},
	month = may,
	year = {2001},
	pages = {912--914},
}

@article{ovchinnikov_diffraction_1999,
	title = {Diffraction of a Released Bose-Einstein Condensate by a Pulsed Standing Light Wave},
	volume = {83},
	issn = {0031-9007, 1079-7114},
	url = {https://link.aps.org/doi/10.1103/PhysRevLett.83.284},
	doi = {10.1103/PhysRevLett.83.284},
	language = {en},
	number = {2},
	urldate = {2025-12-08},
	journal = {Phys. Rev. Lett.},
	author = {Ovchinnikov, Yu. B. and M{\"u}ller, J. H. and Doery, M. R. and Vredenbregt, E. J. D. and Helmerson, K. and Rolston, S. L. and Phillips, W. D.},
	month = jul,
	year = {1999},
	pages = {284--287},
}

@article{mair_entanglement_2001,
	title = {Entanglement of the orbital angular momentum states of photons},
	volume = {412},
	issn = {1476-4687},
	url = {https://doi.org/10.1038/35085529},
	doi = {10.1038/35085529},
	number = {6844},
	journal = {Nature},
	author = {Mair, Alois and Vaziri, Alipasha and Weihs, Gregor and Zeilinger, Anton},
	month = jul,
	year = {2001},
	pages = {313--316},
}

@article{fickler_quantum_2012,
	title = {Quantum Entanglement of High Angular Momenta},
	volume = {338},
	issn = {0036-8075, 1095-9203},
	url = {https://www.science.org/doi/10.1126/science.1227193},
	doi = {10.1126/science.1227193},
	language = {en},
	number = {6107},
	urldate = {2025-12-08},
	journal = {Science},
	author = {Fickler, Robert and Lapkiewicz, Radek and Plick, William N. and Krenn, Mario and Schaeff, Christoph and Ramelow, Sven and Zeilinger, Anton},
	month = nov,
	year = {2012},
	pages = {640--643},
}

@article{willig_sted_2006,
	title = {{STED} microscopy reveals that synaptotagmin remains clustered after synaptic vesicle exocytosis},
	volume = {440},
	issn = {0028-0836, 1476-4687},
	url = {https://www.nature.com/articles/nature04592},
	doi = {10.1038/nature04592},
	language = {en},
	number = {7086},
	urldate = {2025-12-08},
	journal = {Nature},
	author = {Willig, Katrin I. and Rizzoli, Silvio O. and Westphal, Volker and Jahn, Reinhard and Hell, Stefan W.},
	month = apr,
	year = {2006},
	pages = {935--939},
}

@article{allen_orbital_1992,
	title = {Orbital angular momentum of light and the transformation of {Laguerre}-{Gaussian} laser modes},
	volume = {45},
	copyright = {http://link.aps.org/licenses/aps-default-license},
	issn = {1050-2947, 1094-1622},
	url = {https://link.aps.org/doi/10.1103/PhysRevA.45.8185},
	doi = {10.1103/PhysRevA.45.8185},
	language = {en},
	number = {11},
	urldate = {2026-01-13},
	journal = {Phys. Rev. A},
	author = {Allen, L. and Beijersbergen, M. W. and Spreeuw, R. J. C. and Woerdman, J. P.},
	month = jun,
	year = {1992},
	pages = {8185--8189},
}

@article{milosevic_above-threshold_2006,
	title = {Above-threshold ionization by few-cycle pulses},
	volume = {39},
	issn = {0953-4075, 1361-6455},
	url = {https://iopscience.iop.org/article/10.1088/0953-4075/39/14/R01},
	doi = {10.1088/0953-4075/39/14/R01},
	number = {14},
	urldate = {2026-01-23},
	journal = {J. Phys. B: At. Mol. Opt. Phys.},
	author = {Milo{\v s}evi{\'c}, D B and Paulus, G G and Bauer, D and Becker, W},
	month = jul,
	year = {2006},
	pages = {R203--R262},
}

@article{de_ninno_photoelectric_2020,
	title = {Photoelectric effect with a twist},
	volume = {14},
	issn = {1749-4885, 1749-4893},
	url = {https://www.nature.com/articles/s41566-020-0669-y},
	doi = {10.1038/s41566-020-0669-y},
	language = {en},
	number = {9},
	urldate = {2026-02-25},
	journal = {Nat. Photon.},
	author = {De Ninno, Giovanni and W{\"a}tzel, Jonas and Ribi{\v c}, Primo{\v z} Rebernik and Allaria, Enrico and Coreno, Marcello and Danailov, Miltcho B. and David, Christian 
	and Demidovich, Alexander and Di Fraia, Michele and Giannessi, Luca and Hansen, Klavs and Kru{\v s}i{\v c}, {\v S}pela and Manfredda, Michele and Meyer, Michael and Miheli{\v c}, 
	Andrej and Mirian, Najmeh and Plekan, Oksana and Ressel, Barbara and R{\"o}sner, Benedikt and Simoncig, Alberto and Spampinati, Simone and Stupar, Matija and {\v Z}itnik, Matja{\v z} and 
	Zangrando, Marco and Callegari, Carlo and Berakdar, Jamal},
	month = sep,
	year = {2020},
	pages = {554--558},

}

@article{fang_photoelectronic_2021,
	title = {Photoelectronic mapping of the spin–orbit interaction of intense light fields},
	volume = {15},
	issn = {1749-4885, 1749-4893},
	url = {https://www.nature.com/articles/s41566-020-00709-3},
	doi = {10.1038/s41566-020-00709-3},
	language = {en},
	number = {2},
	urldate = {2026-02-25},
	journal = {Nat. Photon.},
	author = {Fang, Yiqi and Han, Meng and Ge, Peipei and Guo, Zhenning and Yu, Xiaoyang and Deng, Yongkai and Wu, Chengyin and Gong, Qihuang and Liu, Yunquan},
	month = feb,
	year = {2021},
	pages = {115--120},
}

@article{fang_orbital_2020,
	title = {Orbital angular momentum holography for high-security encryption},
	volume = {14},
	issn = {1749-4885, 1749-4893},
	url = {https://www.nature.com/articles/s41566-019-0560-x},
	doi = {10.1038/s41566-019-0560-x},
	language = {en},
	number = {2},
	urldate = {2026-03-09},
	journal = {Nat. Photon.},
	author = {Fang, Xinyuan and Ren, Haoran and Gu, Min},
	month = feb,
	year = {2020},
	pages = {102--108},
}

@article{gu_dynamical_2026,
	title = {Dynamical Phase Evolution of Coulomb-Focused Electrons in Strong-Field Ionization Probed by a Standing Light Wave},
	volume = {136},
	issn = {0031-9007, 1079-7114},
	url = {https://link.aps.org/doi/10.1103/63v1-3b63},
	doi = {10.1103/63v1-3b63},
	number = {10},
	urldate = {2026-03-12},
	journal = {Phys. Rev. Lett.},
	author = {Gu, Yuan and Liang, Hao and Zheng, Weiran and Lin, Aofan and Zhang, Jiaye and Li, Zichen and Du, Juan and Ying, Lei and He, Peilun and Rost, Jan-Michael and Jacob, Sina and Kunitski,
	 Maksim and Jahnke, Till and Eckart, Sebastian and Lin, Kang and D{\"o}rner, Reinhard},
	month = mar,
	year = {2026},
	pages = {103201},
}

@article{blaga_strong-field_2009,
	title = {Strong-field photoionization revisited},
	volume = {5},
	issn = {1745-2473, 1745-2481},
	url = {https://www.nature.com/articles/nphys1228},
	doi = {10.1038/nphys1228},
	number = {5},
	urldate = {2026-04-24},
	journal = {Nat. Phys.},
	author = {Blaga, C. I. and Catoire, F. and Colosimo, P. and Paulus, G. G. and Muller, H. G. and Agostini, P. and DiMauro, L. F.},
	month = may,
	year = {2009},
	pages = {335--338},
}

@article{freimund_bragg_2002,
	title = {Bragg Scattering of Free Electrons Using the Kapitza-Dirac Effect},
	volume = {89},
	copyright = {http://link.aps.org/licenses/aps-default-license},
	issn = {0031-9007, 1079-7114},
	url = {https://link.aps.org/doi/10.1103/PhysRevLett.89.283602},
	doi = {10.1103/PhysRevLett.89.283602},
	language = {en},
	number = {28},
	urldate = {2026-04-24},
	journal = {Phys. Rev. Lett.},
	author = {Freimund, Daniel L. and Batelaan, Herman},
	month = dec,
	year = {2002},
	pages = {283602},
}

@article{fang_ultrafast_2025,
	title = {Ultrafast physics with structured light},
	volume = {7},
	issn = {2522-5820},
	url = {https://www.nature.com/articles/s42254-025-00887-5},
	doi = {10.1038/s42254-025-00887-5},
	language = {en},
	number = {12},
	urldate = {2026-04-24},
	journal = {Nat. Rev. Phys.},
	author = {Fang, Yiqi and Lyu, Zijian and Liu, Yunquan},
	month = nov,
	year = {2025},
	pages = {713--727},
}

@article{zuo_non-cascade_2026,
	title = {Non-cascade random walks in solid-state high harmonic generation},
	volume = {17},
	issn = {2041-1723},
	url = {https://www.nature.com/articles/s41467-026-69668-7},
	doi = {10.1038/s41467-026-69668-7},
	language = {en},
	number = {1},
	urldate = {2026-04-27},
	journal = {Nat. Commun.},
	author = {Zuo, Zitan and Wang, Yiwen and Pan, Shengzhe and Han, Lulu and Xu, Yidan and Wu, Dian and Jiang, Shicheng and Wu, Jian},
	month = feb,
	year = {2026},
	pages = {2912},
}

@article{fang_probing_2022,
	title = {Probing the orbital angular momentum of intense vortex pulses with strong-field ionization},
	volume = {11},
	issn = {2047-7538},
	url = {https://www.nature.com/articles/s41377-022-00726-7},
	doi = {10.1038/s41377-022-00726-7},
	language = {en},
	number = {1},
	urldate = {2026-04-27},
	journal = {Light Sci. Appl.},
	author = {Fang, Yiqi and Guo, Zhenning and Ge, Peipei and Dou, Yankun and Deng, Yongkai and Gong, Qihuang and Liu, Yunquan},
	month = feb,
	year = {2022},
	pages = {34},
}

@article{schmiegelow_transfer_2016,
	title = {Transfer of optical orbital angular momentum to a bound electron},
	volume = {7},
	issn = {2041-1723},
	url = {https://www.nature.com/articles/ncomms12998},
	doi = {10.1038/ncomms12998},
	number = {1},
	urldate = {2026-04-27},
	journal = {Nat. Commun.},
	author = {Schmiegelow, Christian T. and Schulz, Jonas and Kaufmann, Henning and Ruster, Thomas and Poschinger, Ulrich G. and Schmidt-Kaler, Ferdinand},
	month = oct,
	year = {2016},
	pages = {12998},
}

@article{xu_molecular_2025,
	title = {Molecular Wave Plate for the Control of Ultrashort Pulses Carrying Orbital Angular Momentum},
	volume = {135},
	issn = {0031-9007, 1079-7114},
	url = {https://link.aps.org/doi/10.1103/j2jj-1jns},
	doi = {10.1103/j2jj-1jns},
	number = {11},
	urldate = {2026-04-27},
	journal = {Phys. Rev. Lett.},
	author = {Xu, Chengqing and He, Lixin and Tao, Wanchen and Zhu, Xiaosong and Wang, Feng and Xu, Long and Xu, Lu and Lan, Pengfei and Averbukh, Ilya and Prior, Yehiam and Lu, Peixiang},
	month = sep,
	year = {2025},
	pages = {113201},
}

@article{begin_orbital_2025,
	title = {Orbital angular momentum control of strong-field ionization in atoms and molecules},
	volume = {16},
	issn = {2041-1723},
	url = {https://doi.org/10.1038/s41467-025-57618-8},
	doi = {10.1038/s41467-025-57618-8},
	number = {1},
	journal = {Nat. Commun.},
	author = {B{\'e}gin, Jean-Luc and Karimi, Ebrahim and Corkum, Paul and Brabec, Thomas and Bhardwaj, Ravi},
	month = mar,
	year = {2025},
	pages = {2467},
}

@article{li_attosecond_2026,
	title = {Attosecond Vortex Photoelectron Holography for Probing Phase-Encoded Chirality},
	volume = {136},
	url = {https://link.aps.org/doi/10.1103/gcxt-18gk},
	doi = {10.1103/gcxt-18gk},
	number = {9},
	journal = {Phys. Rev. Lett.},
	publisher = {American Physical Society},
	author = {Li, Liding and Chen, Yongkun and Yu, Miao and Zhang, Xu and Li, Yang and Zhou, Yueming and Lu, Peixiang},
	month = mar,
	year = {2026},
	pages = {093202},
}

@article{kong_controlling_2017,
	title = {Controlling the orbital angular momentum of high harmonic vortices},
	volume = {8},
	issn = {2041-1723},
	url = {https://doi.org/10.1038/ncomms14970},
	doi = {10.1038/ncomms14970},
	number = {1},
	journal = {Nat. Commun.},
	author = {Kong, Fanqi and Zhang, Chunmei and Bouchard, Fr{\'e}d{\'e}ric and Li, Zhengyan and Brown, Graham G. and Ko, Dong Hyuk and Hammond, T. J. and Arissian, Ladan and Boyd, Robert W. and Karimi, Ebrahim and Corkum, P. B.},
	month = apr,
	year = {2017},
	pages = {14970},
}

@article{zurch_strong-field_2012,
	title = {Strong-field physics with singular light beams},
	volume = {8},
	issn = {1745-2481},
	url = {https://doi.org/10.1038/nphys2397},
	doi = {10.1038/nphys2397},
	number = {10},
	journal = {Nat. Phys.},
	author = {Z{\"u}rch, M. and Kern, C. and Hansinger, P. and Dreischuh, A. and Spielmann, Ch.},
	month = oct,
	year = {2012},
	pages = {743--746},
}

@article{dorney_controlling_2019,
	title = {Controlling the polarization and vortex charge of attosecond high-harmonic beams via simultaneous spin–orbit momentum conservation},
	volume = {13},
	issn = {1749-4893},
	url = {https://doi.org/10.1038/s41566-018-0304-3},
	doi = {10.1038/s41566-018-0304-3},
	number = {2},
	journal = {Nat. Photon.},
	author = {Dorney, Kevin M. and Rego, Laura and Brooks, Nathan J. and San Rom{\'a}n, Julio and Liao, Chen-Ting and Ellis, Jennifer L. and Zusin, 
	Dmitriy and Gentry, Christian and Nguyen, Quynh L. and Shaw, Justin M. and Pic{\'o}n, Antonio and Plaja, Luis and Kapteyn, Henry C. and Murnane, Margaret M. and Hern{\'a}ndez-Garc{\'i}a, Carlos},
	month = feb,
	year = {2019},
	pages = {123--130},
}

@article{gariepy_creating_2014,
	title = {Creating High-Harmonic Beams with Controlled Orbital Angular Momentum},
	volume = {113},
	url = {https://link.aps.org/doi/10.1103/PhysRevLett.113.153901},
	doi = {10.1103/PhysRevLett.113.153901},
	number = {15},
	journal = {Phys. Rev. Lett.},
	publisher = {American Physical Society},
	author = {Gariepy, Genevieve and Leach, Jonathan and Kim, Kyung Taec and Hammond, T. J. and Frumker, E. and Boyd, Robert W. and Corkum, P. B.},
	month = oct,
	year = {2014},
	pages = {153901},
}

@article{rego_generation_2019,
	title = {Generation of extreme-ultraviolet beams with time-varying orbital angular momentum},
	volume = {364},
	url = {https://www.science.org/doi/abs/10.1126/science.aaw9486},
	doi = {10.1126/science.aaw9486},
	number = {6447},
	journal = {Science},
	author = {Rego, Laura and Dorney, Kevin M. and Brooks, Nathan J. and Nguyen, Quynh L. and Liao, Chen-Ting and Rom{\'a}n, Julio San and Couch, David E. and Liu, Allison and Pisanty, Emilio and Lewenstein, Maciej and Plaja, Luis and Kapteyn, Henry C. and Murnane, Margaret M. and Hern{\'a}ndez-Garc{\'i}a, Carlos},
	year = {2019},
	pages = {eaaw9486},
}

@article{gui_second-harmonic_2021,
	title = {Second-harmonic generation and the conservation of spatiotemporal orbital angular momentum of light},
	volume = {15},
	issn = {1749-4893},
	url = {https://doi.org/10.1038/s41566-021-00841-8},
	doi = {10.1038/s41566-021-00841-8},
	number = {8},
	journal = {Nat. Photon.},
	author = {Gui, Guan and Brooks, Nathan J. and Kapteyn, Henry C. and Murnane, Margaret M. and Liao, Chen-Ting},
	month = aug,
	year = {2021},
	pages = {608--613},
}

@article{hancock_second-harmonic_2021,
	title = {Second-harmonic generation of spatiotemporal optical vortices and conservation of orbital angular momentum},
	volume = {8},
	url = {https://opg.optica.org/optica/abstract.cfm?URI=optica-8-5-594},
	doi = {10.1364/OPTICA.422743},
	number = {5},
	journal = {Optica},
	publisher = {Optica Publishing Group},
	author = {Hancock, S. W. and Zahedpour, S. and Milchberg, H. M.},
	month = may,
	year = {2021},
	pages = {594--597},
}

@article{Keldysh_ionization_1965,
	title = {Ionization in the field of a strong electromagnetic wave},
	volume = {20},
	pages = {1307},
	journal = {Sov. Phy. JETP},
	author = {Keldysh, L. V.},
	month = nov,
	year = {1965},
}

\end{document}